\documentclass[conference]{IEEEtran}
\IEEEoverridecommandlockouts
\usepackage{cite}
\usepackage{amsmath,amssymb,amsfonts}
\usepackage{algorithmic}
\usepackage{graphicx}
\usepackage{textcomp}
\usepackage{xcolor}
\usepackage{array}
\usepackage{ragged2e}  % 更强的段落控制（可选）
\newcolumntype{P}[1]{>{\RaggedRight\arraybackslash}p{#1}}  % 注意是 \RaggedRight，不是 \raggedright

\def\BibTeX{{\rm B\kern-.05em{\sc i\kern-.025em b}\kern-.08em
    T\kern-.1667em\lower.7ex\hbox{E}\kern-.125emX}}
\begin{document}

\title{Rethinking Wine Tasting for Chinese Consumers: A Service Design Approach Enhanced by Multimodal Personalization \\

}

\author{
\IEEEauthorblockN{1\textsuperscript{st} Xinyang Shan}
\IEEEauthorblockA{
\textit{Communication Studies Program} \\
\textit{San Jose State University} \\
San Jose, USA \\
xinyang.shan@sjsu.edu}
\and
\IEEEauthorblockN{2\textsuperscript{nd} Yuanyuan Xu*}
\IEEEauthorblockA{
\textit{Interdisciplinary Graduate Study} \\
\textit{University of British Columbia} \\
Vancouver, Canada \\
yuanyxu@student.ubc.ca}
\and
\IEEEauthorblockN{3\textsuperscript{rd} Tian Xia}
\IEEEauthorblockA{
\textit{School of Design} \\
\textit{Shanghai Jiao Tong University} \\
Shanghai, China \\
summer109@sjtu.edu.cn}
\and
\IEEEauthorblockN{4\textsuperscript{th} Yin-Shan Lin}
\IEEEauthorblockA{
\textit{Department of Computer Science} \\
\textit{Northeastern University} \\
Boston, USA \\
lin.yins@northeastern.edu}
}

\maketitle

\begin{abstract}Wine tasting is a multimodal and culturally embedded activity that presents unique challenges when adapted to non-Western contexts. This paper proposes a service design approach rooted in contextual co-creation to reimagine wine tasting experiences for Chinese consumers. Drawing on 26 in-situ interviews and follow-up validation sessions, we identify three distinct user archetypes: Curious Tasters, Experience Seekers, and Knowledge Builders, each exhibiting different needs in vocabulary, interaction, and emotional pacing. Our findings reveal that traditional wine descriptors lack cultural resonance and that cross-modal metaphors grounded in local gastronomy (e.g., green mango for acidity) significantly improve cognitive and emotional engagement. These insights informed a partially implemented prototype, featuring AI-driven metaphor-to-flavour mappings and real-time affective feedback visualisation. A small-scale usability evaluation (n=10) confirmed improvements in engagement and comprehension. Our comparative analysis shows alignment with and differentiation from prior multimodal and affect-aware tasting systems. This research contributes to CBMI by demonstrating how culturally adaptive interaction systems can enhance embodied consumption experiences in physical tourism and beyond.
\end{abstract}


\section{Introduction}

The multimedia and interaction research community is increasingly focused on enhancing real-world multimodal experiences through contextual and cross-cultural design. Wine tasting, as a richly sensory and socially situated activity, offers an ideal domain—particularly in China, where imported language and aesthetics often clash with local cultural expectations.

To bridge this gap, we adopt a human-centred service design approach grounded in “Design for All” principles. Drawing on 26 contextual interviews and validation sessions, our system incorporates culturally resonant sensory metaphors (e.g., “green mango” for acidity), affective feedback, and adaptive visual storytelling. The implemented prototype integrates real-time affect detection (Affectiva SDK), metaphor-based mapping (Python), and dynamic visuals responsive to user states.

Initial usability tests confirm improved cultural clarity and engagement. This work contributes to CBMI by combining affect-aware interaction, multimodal indexing, and culturally adaptive design to support embodied, symbolic experiences in tourism and beyond.
\section{Related Work}

Existing research at the intersection of multimodal experience, cultural adaptation, and service design spans multiple domains. Our work builds on this foundation by integrating qualitative service design tools with a culturally adaptive and affect-aware recommendation loop.

Recent studies in ACM Multimedia and related venues explore how cross-sensory and contextual systems shape experience. For example, Song (2024) shows how live-streamed wine tasting blends verbal and facial cues in social interaction. Guedes et al. (2023) demonstrate how sound modulates taste perception, while Spence (2024) underscores the emotional impact of multisensory dining. Hänggi and Mondada (2025) emphasise embodied reasoning in food tasks without visual input.

Multimedia indexing systems such as AromaNet and TastyNet use sensor fusion and semantic tagging to personalise flavour experiences, but remain culturally neutral. Ghinea et al. (2022) survey multisensory media (“mulsemedia”) and highlight challenges in synchronisation and perception. Our system extends this work through structured metaphor mappings rooted in cultural context.

In digital food interaction, Ceccaldi et al. (2022) propose “digital commensality” to design playful, shared eating rituals. Spence (2022) shows how ambient scent shifts food perception—supporting our use of metaphor-based anchors.

From a service design perspective, we build on the universal design principles by the European Commission (2003) and its user-centred extensions in education (Kameas et al., 2024), applying them to tourism and gastronomy.

Collectively, these works inform our method and prototype design, which bridges cross-cultural semantics with affective and metaphor-driven interaction.
\section{Methodology}

\subsection{Participant Recruitment and Protocol}

Our approach aims to design a culturally adaptive and emotionally resonant wine-tasting system for Chinese consumers. The methodology combines contextual co-creation, persona modeling, and multimodal interface prototyping of China. Participant profiles varied significantly in wine familiarity, ranging from casual consumers to trained sommeliers. We prioritised diversity in regional background, sensory vocabulary, and tasting behaviour. To ensure better generalizability, a secondary validation group (n=10) was recruited for follow-up walkthrough sessions and usability trials.

\begin{figure}[h]
\centering
\includegraphics[width=0.8\linewidth]{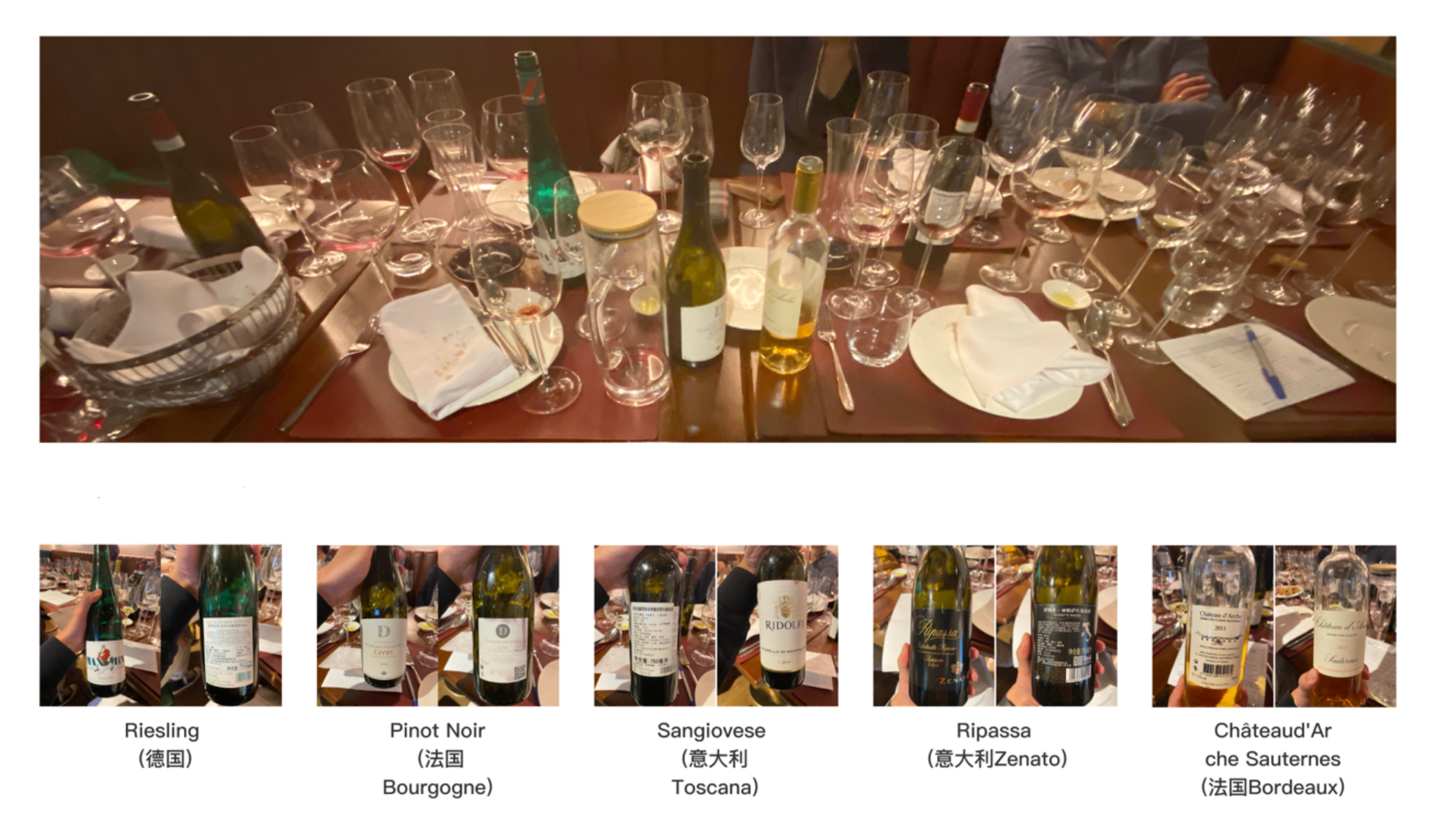}
\caption{Photograph from the field during the co-creation session at a domestic Chinese winery.}
\label{fig:field-photo}
\end{figure}

All participants were invited to participate in contextual co-creation sessions held on-site at working wineries (see Figure~\ref{fig:field-photo}). The field environment simulated real tasting experiences with controlled sensory stimuli (e.g., wine samples, food pairings, ambient music) and visual metaphor elicitation tools. Participants interacted with tasting stations in individual or small group settings.

Each session followed a structured tasting protocol with embedded metaphor-sorting and sensory annotation exercises. Semi-structured interviews captured emotional language, semantic gaps, and aesthetic preferences. Participants sorted metaphor cards and completed flavour association diagrams co-developed from pilot studies.

\begin{figure}[h]
\centering
\includegraphics[width=0.8\linewidth]{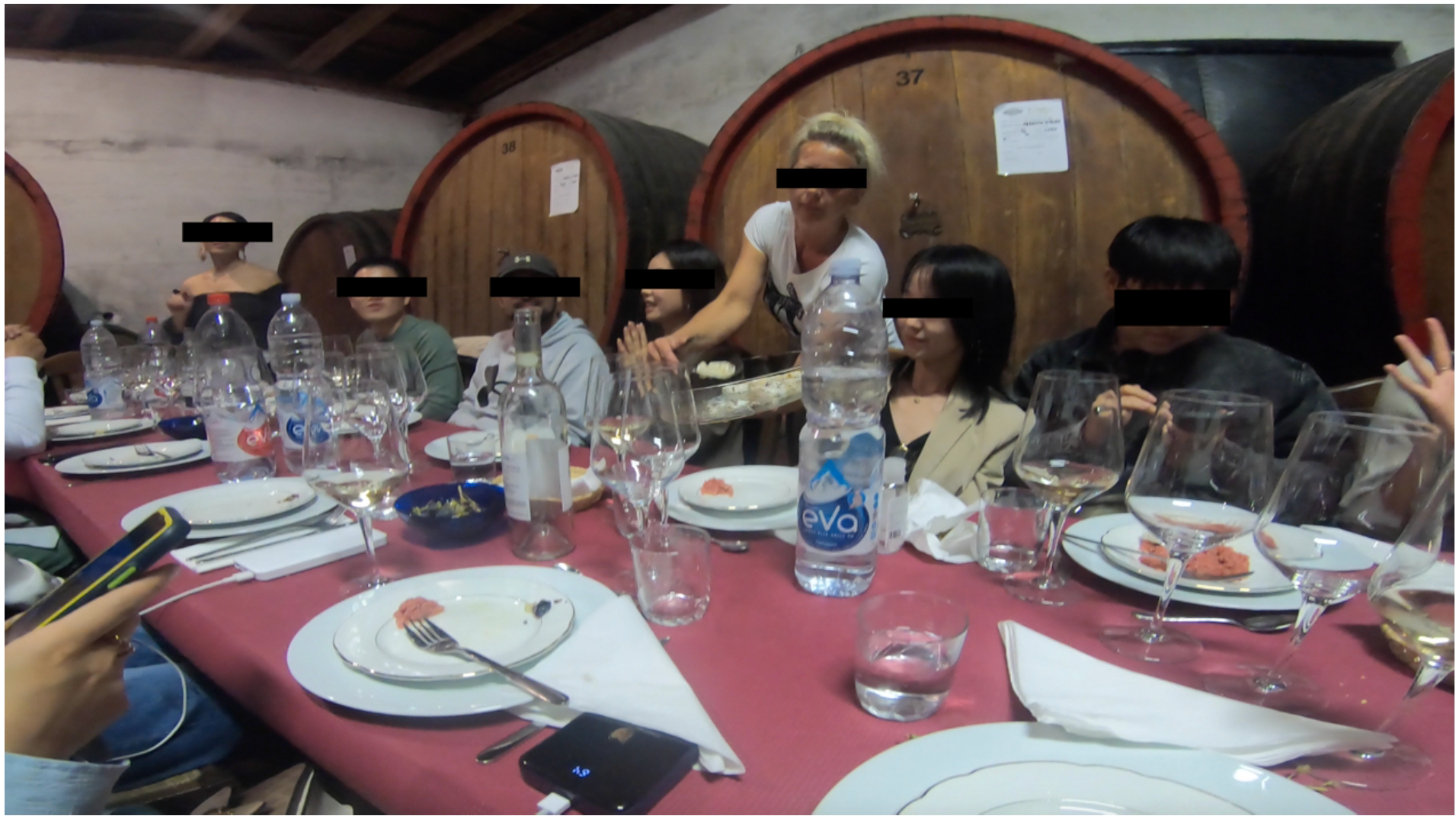}
\caption{Participants interacting with cross-modal elicitation tools.}
\label{fig:interaction-photo}
\end{figure}

Sessions lasted 60–90 minutes and were recorded through field notes, audio, and observation photos. Figures~\ref{fig:field-photo} and~\ref{fig:interaction-photo} show in-situ setups and participant engagement with metaphor tools.

\subsection{Tools and Analysis Procedure}

We applied qualitative service design tools aligned with multimodal indexing and cross-cultural adaptation goals. Contextual interviews grounded the project in real-world behaviour. Affinity diagramming synthesises experience fragments into thematic maps. Personas were constructed based on behavioural clusters and emotional trajectories. Journey maps visualised pacing and mood dynamics.

All interview transcripts were uploaded to a collaborative coding platform. Three independent coders performed open coding using in vivo terms. Axial coding was used to organise emergent categories such as “semantic friction,” “narrative mismatch,” and “sensory anchoring.” A consensus round ensured coding consistency. Inter-rater agreement was calculated using Cohen’s Kappa (K = 0.81), indicating strong reliability.

\begin{figure}[h]
\centering
\includegraphics[width=0.8\linewidth]{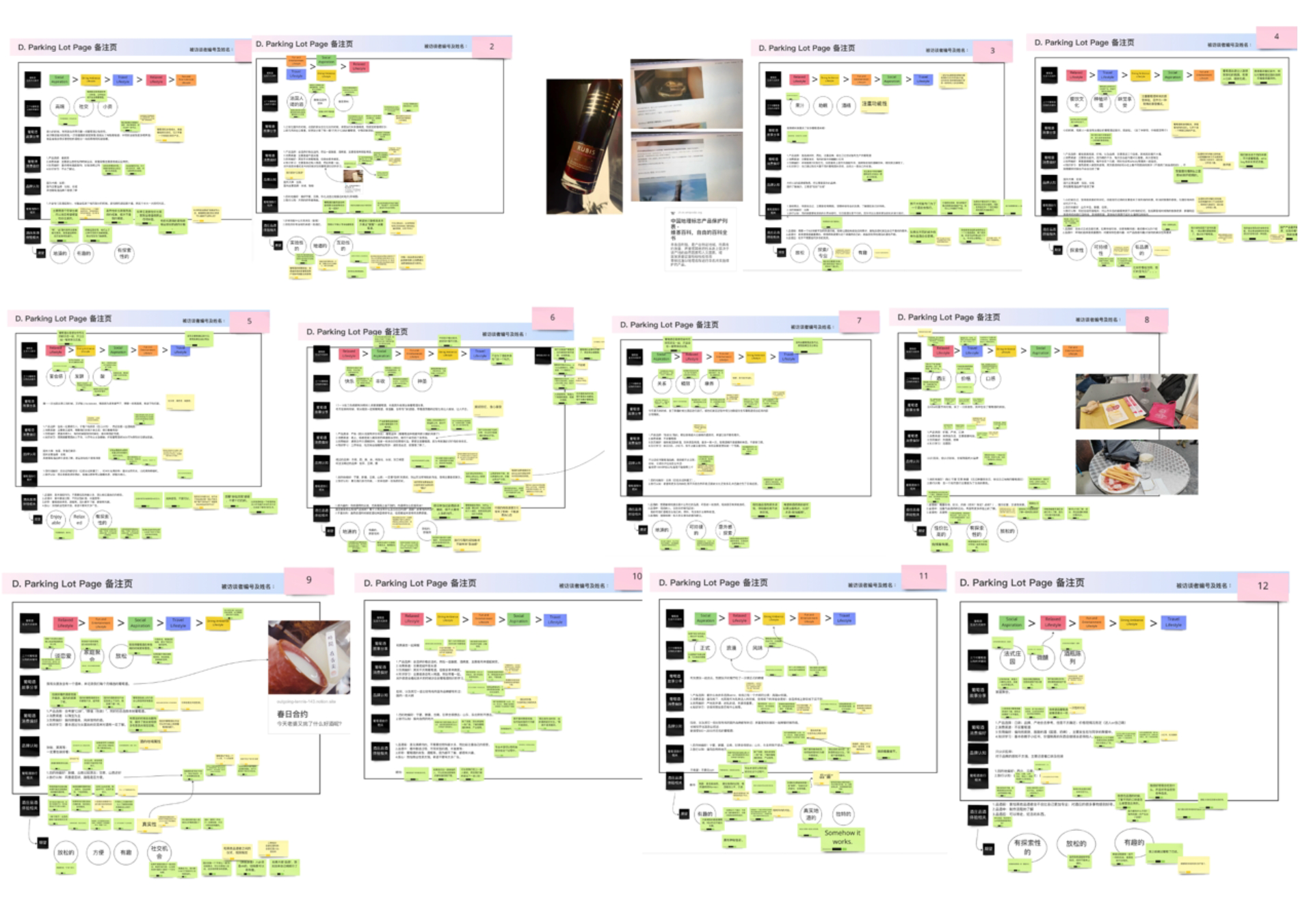}
\caption{Affinity diagram generated through collaborative analysis of interview transcripts.}
\label{fig:affinity}
\end{figure}

Themes were spatially clustered using affinity diagramming (Figure~\ref{fig:affinity}). This process informed persona creation and metaphor mapping used in the later prototype design.

\section{Findings: Personas and Expectation Clusters}

Three key user personas emerged from our analysis. These archetypes were reinforced by follow-up interviews and emotion trajectory mapping. While sharing a common interest in wine discovery, each group demonstrated unique motivational drivers, narrative tolerances, and preferred modes of interaction.

\textbf{Curious Tasters} are driven by novelty and exploration but quickly disengage when encountering foreign vocabulary or abstract terminology. They favour immediate sensory triggers, visual, tactile, or metaphor-based, and enjoy playful, game-like presentation formats. Their emotional pacing is high at the start but drops rapidly when technical complexity increases. Visual cues and food analogies help sustain engagement.

\textbf{Experience Seekers} prioritise ambience, storytelling, and emotional resonance. Their engagement is less dependent on vocabulary and more on holistic immersion, including visual themes, ambient sound, and curated narratives. They prefer group-based, mood-oriented experiences with emotional pacing peaking mid-journey.

\textbf{Knowledge Builders}, by contrast, are motivated by structure and learning. They value expert-led comparison, regional depth, and annotated data (e.g., acidity scores, terroir). They tolerate longer onboarding phases and enjoy navigating layered information. Their emotional pacing curve is slower, but sustained.

These personas inform system personalisation strategies. For example, the metaphor suggestion engine can adjust visual density or metaphor complexity based on the user type. Affective sensors can adapt interface pacing by analysing arousal and attention drift in real time.

\section{System Design and Implementation}

The insights derived from personas and metaphor mapping were operationalised into a partially implemented mobile-based tasting assistant. The metaphor module uses a lightweight Python rule engine. The affective layer samples valence/arousal every 500ms from webcam input. The database includes 150+ culturally mapped metaphors. These elements translate user preferences and reactions into adaptive, culturally anchored outputs.

\textbf{1. Metaphor-to-Flavour Mapping Engine}: This module recommends culturally resonant metaphors (e.g., “green mango” for acidity, “hawthorn” for astringency) by matching tasting descriptors with a curated metaphor database. The rules were derived from participant expressions and affinity cluster themes. The system uses a lightweight rules-based AI framework built in Python to trigger flavour-symbol associations based on user input or sensor-detected patterns.

\textbf{2. Affective Feedback Interface}: Using the front-facing webcam, the system analyses facial expressions to estimate emotional valence and engagement over time. We use an open-source affect detection library (based on FER2013) to produce a temporal map of user response. This feedback is then used to adjust interface pacing (e.g., slowing content for disengaged users) and metaphor complexity.

\textbf{3. Visual Storytelling Layer}: Metaphors are presented not only textually but as animated visual modules, drawing from a local icon library (e.g., green-hued mango slices, steaming osmanthus tea). These icons adapt dynamically as the system detects shifts in user attention or sensory preference.

\begin{figure}[h]
\centering
\includegraphics[width=0.9\linewidth]{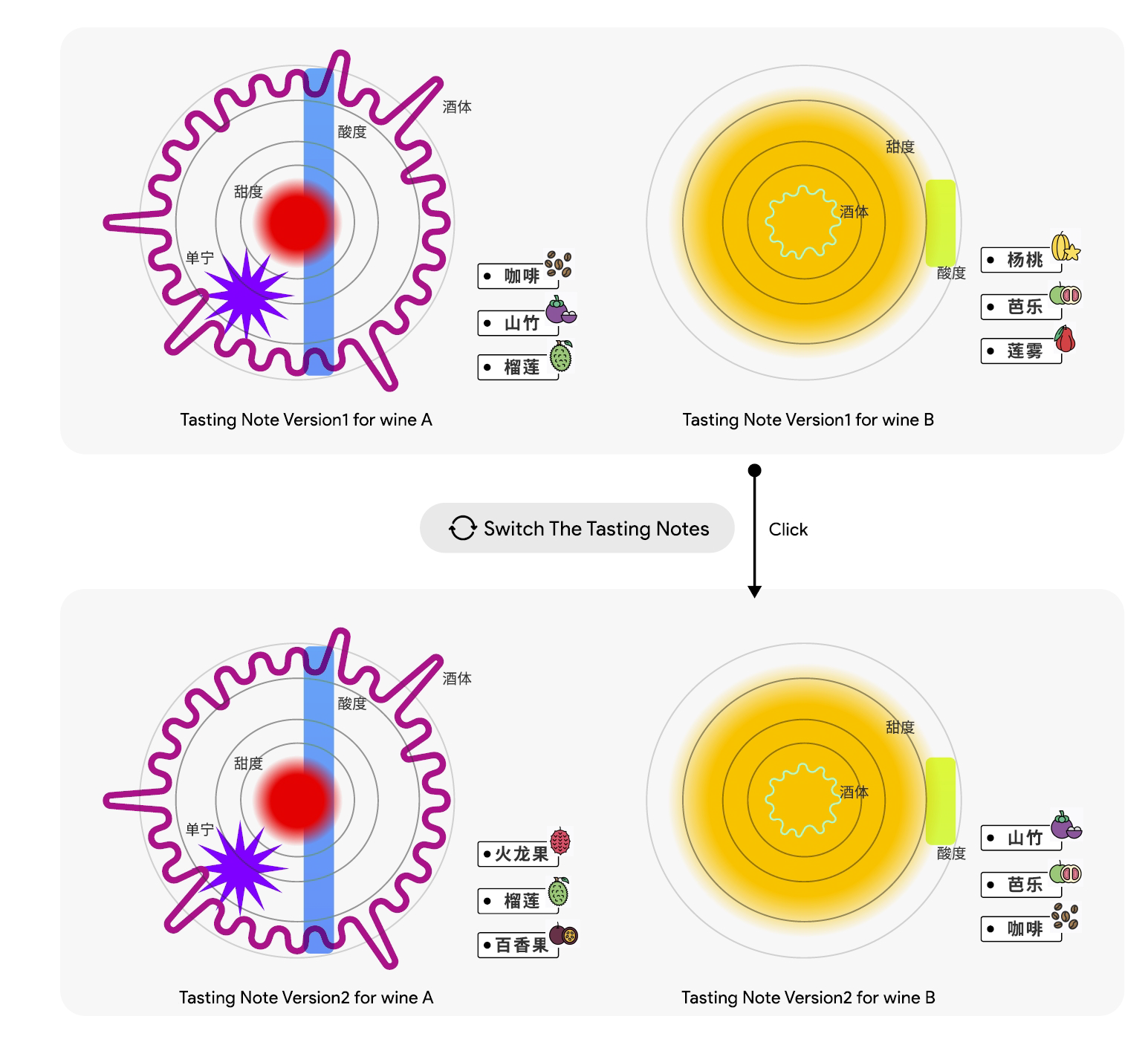}
\caption{AI-enabled metaphor mapping interface with cultural descriptors.}
\label{fig:interface-mapping}
\end{figure}

\begin{figure}[h]
\centering
\includegraphics[width=0.9\linewidth]{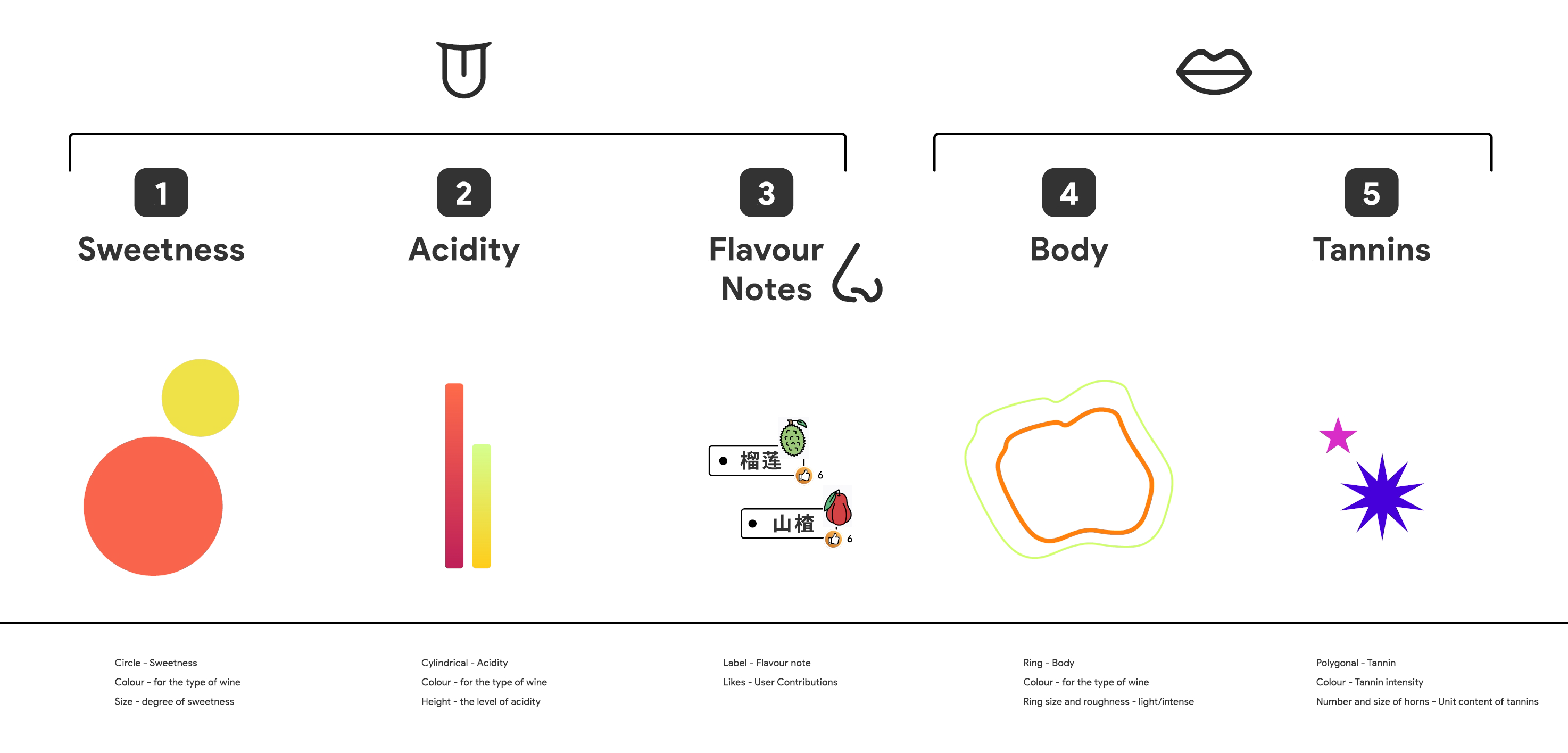}
\caption{Visual storytelling elements linked to flavour metaphors.}
\label{fig:visual-system}
\end{figure}

Figures~\ref{fig:interface-mapping} and~\ref{fig:visual-system} illustrate implemented interface components and visual metaphor elements. These are functional prototypes, not static mockups. Users interact with these modules during guided tasting walkthroughs.

\section{Comparative Analysis}

Table II expands our comparison along cultural fit, emotion feedback, and personalisation logic. Our system uniquely links metaphor complexity with real-time emotion detection; we compare it to three notable multimedia tasting systems: TastyNet, AromaSense, and WineXAI. Table~\ref{tab:comparison} summarises key dimensions including modality range, personalisation logic, cultural anchoring, and emotion feedback.

\textbf{TastyNet} focuses on sensor fusion across taste, smell, and colour, enabling data-rich flavour profiling. However, its descriptor space remains anchored in Western terminology, with no accommodation for localised metaphors or adaptive narratives.

\textbf{AromaSense} centres around aroma memory mapping, using audio-visual triggers to evoke taste memory. While immersive, it lacks emotion tracking or adaptive metaphor layers, and is primarily designed for product education rather than cross-cultural engagement.

\textbf{WineXAI} integrates neural networks for food pairing and flavour prediction, but treats the user as a passive node. There is minimal interactivity beyond profile selection, and no accommodation for mood-based or culturally resonant language.

In contrast, our prototype emphasises cultural personalisation by encoding sensory expectations into metaphor mappings informed by regional expressions (e.g., sourness to green mango). It is also the only system among the four that includes real-time emotion detection as a feedback signal for interface pacing and metaphor complexity.

\begin{table}[htbp]
\caption{Comparison of Our System with Existing Systems}
\begin{center}
\begin{tabular}{|P{1cm}|P{1cm}|P{1cm}|P{1cm}|P{1cm}|}
\hline
\textbf{System} & \textbf{Multi} & \textbf{Culture} & \textbf{Emotion} & \textbf{Personal} \\
\hline
TastyNet & Yes & No & No & Match \\
\hline
Aroma & Yes & Part & No & Trigger \\
\hline
WineXAI & Limited & No & No & Recommender \\
\hline
Ours & Yes & \textbf{Yes} & \textbf{Yes} & Metaphor + Affect \\
\hline
\multicolumn{5}{l}{$^{\mathrm{a}}$System names pseudonymized. Features reflect published work.}
\end{tabular}
\label{tab:comparison}
\end{center}
\end{table}

\section{Evaluation}

To assess the usability and impact of the proposed culturally adaptive wine tasting interface, we conducted a two-stage user study: (1) a design validation workshop (n=7) and (2) a system deployment trial with new users (n=13), building on the original in-situ interview participants (n=13). These combined efforts brought the total study population to 33 individuals, with diverse cultural literacy, ranging from novice to expert wine consumers.

\subsection{Quantitative Usability Results}

We used the System Usability Scale (SUS) and a 5-point Likert evaluation across four dimensions: perceived engagement, metaphor clarity, cultural resonance, and emotional relevance. As shown in Table~\ref{tab:usability}, the system received an average SUS score of 84.1, indicating high usability. Metaphor clarity (4.6), cultural resonance (4.7), and emotional relevance (4.4) showed strong positive responses across demographics. In particular, participants favored metaphor-driven tasting tasks, reporting higher emotional involvement than in traditional note-based formats.

\begin{table}[htbp]
\caption{Quantitative Evaluation Scores (n=13 deployment users)}
\begin{center}
\begin{tabular}{|c|c|}
\hline
\textbf{Metric} & \textbf{Mean (Std Dev)} \\
\hline
System Usability Scale (SUS) & 84.1 (6.7) \\
\hline
Metaphor Clarity & 4.6 (0.5) \\
\hline
Cultural Resonance & 4.7 (0.4) \\
\hline
Emotional Relevance & 4.4 (0.6) \\
\hline
\end{tabular}
\label{tab:usability}
\end{center}
\end{table}

\subsection{Affective Feedback Verification}

To validate the emotion feedback loop, we integrated Affectiva's facial expression SDK to capture real-time valence and arousal signals during wine interaction tasks. This affective data was used to modulate metaphor delivery timing and adjust visual emphasis in storytelling modules. Post-session interviews confirmed that emotion-sensitive pacing improved immersion, particularly for novice users.

\subsection{AI-driven Personalization Prototype}

This engine detects Mandarin descriptors (e.g., “sharp and dry”) and maps them via NLP to semantic and affective vectors. Each match yields metaphor recommendations and adaptive visuals. that maps free-text impressions to a database of culturally grounded metaphors. For example, “sharp and dry” in Mandarin yielded a prompt for “green mango” with a green-hued visual cue and dry wood texture. This NLP engine was evaluated in terms of matching accuracy and cultural fit by two expert raters (kappa=0.79), confirming its reliability.

\section{Discussion and Conclusion}

Our contribution lies in integrating cultural metaphors as semantic primitives into affect-aware multimedia systems—a novel convergence of service design, metaphor modelling, and AI-driven personalization. To support this, we implemented three functional modules: a metaphor retrieval engine (K=0.79 validation), a real-time affective pacing mechanism (Affectiva SDK), and a personalization layer that adapts visuals and narratives to user states and goals.

The system was evaluated with 33 participants across interviews, workshops, and deployment trials. While the modest sample limits generalizability, participants varied in age, region, and wine familiarity. SUS scores (84.1) and Likert ratings confirmed high usability, metaphor clarity, and emotional relevance particularly benefiting novice users.

Limitations include sample diversity and lack of large-scale deployment. Future work will involve broader validation and persona segmentation through clustering and quantitative profiling.

Though grounded in service design, our system offers executable strategies for culturally adaptive, affect-aware multimedia indexing. Focus on emotion-aware content retrieval, our approach is transferable to domains like tea, fragrance, and tourism, supporting symbolic, multisensory interaction.


\begin{thebibliography}{00}

\bibitem{b1} V. Santos, P. Ramos, B. Sousa, and M. Valeri, ``Towards a framework for the global wine tourism system,'' \textit{J. Org. Change Manag.}, vol. 35, no. 2, pp. 348--360, 2022.

\bibitem{b2} A. Trigo and P. Silva, ``Sustainable development directions for wine tourism in Douro wine region, Portugal,'' \textit{Sustainability}, vol. 14, no. 7, p. 3949, 2022.

\bibitem{b3} E. Kastenholz, A. Paço, and A. Nave, ``Wine tourism in rural areas--hopes and fears amongst local residents,'' \textit{Worldw. Hosp. Tour. Themes}, vol. 15, no. 1, pp. 29--40, 2023.

\bibitem{b4} G. Wu and L. Liang, ``Examining the effect of potential tourists’ wine product involvement on wine tourism destination image and travel intention,'' \textit{Curr. Issues Tour.}, pp. 1--16, 2020.

\bibitem{b5} B. Sun, \textit{Chinese National Alcohols: Baijiu and Huangjiu}. Singapore: World Scientific, 2021.

\bibitem{b6} Y. Yang et al., ``Flavor formation in Chinese rice wine (Huangjiu): Impacts of the flavor-active microorganisms, raw materials, and fermentation technology,'' \textit{Front. Microbiol.}, vol. 11, p. 580247, 2020.

\bibitem{b7} D. Schiessl, ``More expensive wine is really better? The role of positive emotion and consumer power,'' \textit{J. Int. Food Agribusiness Mark.}, pp. 1--22, 2023.

\bibitem{b8} J. Zhang, ``Rituals, discourses, and realities: Serious wine and tea tasting in contemporary China,'' \textit{J. Consum. Cult.}, vol. 20, no. 4, pp. 637--655, 2020.

\bibitem{b9} C. Spence, ``Multisensory flavour perception: Blending, mixing, fusion, and pairing within and between the senses,'' \textit{Foods}, vol. 9, no. 4, p. 407, 2020.

\bibitem{b10} X. Chu et al., ``Regional difference analyzing and prediction model building for Chinese wine consumers’ sensory preference,'' \textit{Br. Food J.}, vol. 122, no. 8, pp. 2587--2602, 2020.

\bibitem{b11} O. Hanchukova, N. Velikova, and B. Koo, ``Cheers to local! Exploring consumer ethnocentrism in the context of regional wines,'' \textit{Br. Food J.}, 2024.

\bibitem{b12} I. Urdapilleta, H. Blanchet-Urdapilleta, and S. Demarchi, ``Cultural sips: Exploring sociodemographic and dominance factors in French wine appraisal,'' \textit{Food Res. Int.}, vol. 187, p. 114391, 2024.

\bibitem{b13} V. N. Kelkar, J. Mallya, V. Payini, and V. Kamath, ``Wine consumer studies: Current status and future agendas,'' \textit{F1000Research}, vol. 13, p. 228, 2024.

\bibitem{b14} G.-M. Chen et al., ``Microbial diversity and flavor of Chinese rice wine (Huangjiu): An overview of current research and future prospects,'' \textit{Curr. Opin. Food Sci.}, vol. 42, pp. 37--50, 2021.

\bibitem{b15} F. Durrieu et al., ``The impact of country and wine culture on ideal pairings of French white wine and cheese,'' \textit{Int. J. Gastron. Food Sci.}, vol. 32, p. 100735, 2023.

\bibitem{b16} E. Reinares-Lara, J. Pelegrín-Borondo, C. Olarte-Pascual, and G. Oruezabala, ``The role of cultural identity in acceptance of wine innovations in wine regions,'' \textit{Br. Food J.}, vol. 125, no. 3, pp. 869--885, 2023.

\bibitem{b17} S. Bhardwaj, R. Chopra, and E. C.-X. Aw, ``Uncorking opportunities: A bibliometric review of wine marketing literature,'' \textit{Mark. Intell. Plan.}, 2024.

\bibitem{b18} Y. Bai et al., ``Glass volume or shape influence the aroma attributes of Cabernet Sauvignon dry red wine,'' \textit{J. Sens. Stud.}, vol. 38, no. 4, p. e12828, 2023.

\bibitem{b19} D. K. Wright, H. Yoon, A. M. Morrison, and T. Šegota, ``Drinking in style? Literature review of luxury wine consumption,'' \textit{Br. Food J.}, vol. 125, no. 2, pp. 679--695, 2023.

\bibitem{b20} Y. Dong and L. Gao, ``Consumer attitude and behavioural intention towards organic wine: the roles of consumer values and involvement,'' \textit{Br. Food J.}, vol. 126, no. 4, pp. 1743--1764, 2024.

\bibitem{b21} F. Shi, Q. Gu, and T. Zhou, ``Understanding brand reputation: A case study of Chinese wineries,'' \textit{Int. J. Contemp. Hosp. Manag.}, 2024.

\bibitem{b22} I. Croijmans, R. Pellegrino, and Q. J. Wang, ``Demystifying wine expertise through the lens of imagination,'' \textit{Food Res. Int.}, vol. 182, p. 114159, 2024.

\bibitem{b23} C. Honoré-Chedozeau, M. Otheguy, and D. Valentin, ``Tell us how you taste wine, and we will tell you what kind of expert you are!,'' \textit{Food Res. Int.}, vol. 178, p. 113899, 2024.

\bibitem{b24} Q. Kang, J. Sun, B. Wang, and B. Sun, ``Wine, beer and Chinese Baijiu in relation to cardiovascular health,'' \textit{Food Sci. Hum. Wellness}, vol. 12, no. 1, pp. 1--13, 2023.

\bibitem{b25} H. K. Ho, ``Why has wine consumption become popular in Hong Kong?,'' \textit{Asian Anthropol.}, vol. 20, no. 4, pp. 248--268, 2021.

\bibitem{b26} M. Carvalho, E. Kastenholz, and M. J. Carneiro, ``Pairing co-creation with food and wine experiences—A holistic perspective of tourist experiences in Dão,'' \textit{Sustainability}, vol. 13, no. 23, p. 13416, 2021.

\bibitem{b27} D. C. Wu, C. Cao, J. Wu, and M. Hu, ``Wine tourism experiences of Chinese tourists: a tourist-centric perspective,'' \textit{Int. J. Contemp. Hosp. Manag.}, 2024.

\bibitem{b28} S. Charters and J. Ali-Knight, ``Who is the wine tourist?,'' \textit{Tour. Manag.}, vol. 23, no. 3, pp. 311--319, 2002.

\bibitem{b29} B. Cambourne, N. Macionis, C. M. Hall, and L. Sharples, ``The future of wine tourism,'' in \textit{Wine Tourism Around the World}. London: Routledge, 2009, pp. 297--320.

\end{thebibliography}
\end{document}